# Ultrathin BaTiO$_3$ templates for multiferroic nanostructures

Xumin Chen[1], Seolun Yang[2], Ji-Hyun Kim[2], Hyung-Do Kim[3], Jae-Sung Kim[2], Geoffrey Rojas[1], Ralph Skomski[1], Haidong Lu[1], Anand Bhattacharya[4], Tiffany Santos[4], Nathan Guisinger[4], Matthias Bode[4], Alexei Gruverman[1], and Axel Enders[1]

[1] Department of Physics & Astronomy, University of Nebraska-Lincoln
  Lincoln, NE 68588, USA

[2] Departments of Physics, Sook-Myung Women's University, Seoul, 140-742, Korea

[3] Beamline division, Pohang Acceleration Laboratory (PAL), Pohang 790-784, Korea

[4] Center for Nanoscale Materials, Argonne National Laboratory, Argonne, Illinois 60439, USA

E-mail: a.enders@me.com

**Abstract.** Structural, electronic and dielectric properties of high-quality ultrathin BaTiO$_3$ films are investigated. The films, which are grown by ozone-assisted molecular beam epitaxy on Nb-doped SrTiO$_3$ (001) substrates and having thicknesses as thin 8 unit cells (3.2 nm), are unreconstructed and atomically smooth with large crystalline terraces. A strain-driven transition to 3D island formation is observed for films of of 13 unit cells thickness (5.2 nm). The high structural quality of the surfaces, together with the dielectric properties similar to bulk BaTiO$_3$ and dominantly TiO$_2$ surface termination, make these films suitable templates for the synthesis of high-quality metal-oxide multiferroic heterostructures for the fundamental study and exploitation of magneto-electric effects, such as a recently proposed interface effect in Fe/BaTiO$_3$ heterostructures based on Fe-Ti interface bonds.

# 1. Introduction

The coexistence of ferroelectricity and ferromagnetism in two-phase multiferroic systems comprising metals and oxides can result in interesting and useful phenomena, such as magneto-electric effects [1]. A model two-phase system exhibiting magneto-electric behavior, that is, the electric field dependence of the magnetization, is epitaxial Fe on $BaTiO_3$ substrates. The magnetic anisotropy of Fe, observable as the magnitude of the coercive field, depends strongly on the $BaTiO_3$'s electric polarization, as has been shown by Sahoo et al. [2]. The underlying mechanism here is a lateral strain exerted by the ferroelectric on the ferromagnet and the associated change of the magneto-elastic contribution to the total magnetic anisotropy. A second mechanism, different in nature but also leading to magneto-electric behavior, has been predicted recently for $Fe/BaTiO_3$ heterostructures [3, 4]. Characteristic for this new effect are modulations of the Fe - Ti bonds at the interface by the piezoelectric distortion of the $BaTiO_3$, resulting in changes of the effective magnetic moments of Fe and Ti atoms and the anisotropy. A second and related example is magnetic tunnel junctions, which are also based on metal-oxide interfaces [5]. Experimental investigation of any metal-oxide heterostructures depends critically on the quality of the oxide layer, since their performance is extremely sensitive to the chemical and structural properties of the interfaces. The realization of model structures or even devices exhibiting and exploiting above effects is complicated by the experimental difficulty to grow metal-oxide interfaces of sufficiently high interface quality.

Atomically smooth $BaTiO_3$ substrates are a requirement for the study of magneto-electric effects in all $BaTiO_3$–based heterostructures. They can be obtained by annealing bulk samples under ultrahigh vacuum or in hydrogen atmosphere to temperatures around 1000 K [6, 7, 8], but both the atomic structure and the surface termination depend critically on the preparation conditions [8, 9, 10]. As alternative to bulk $BaTiO_3$ substrates, ultrathin films of $BaTiO_3$ have been prepared and studied recently [11, 12-27]. Their structural characterization has thus far been limited to reflection high-energy electron diffraction (RHEED) during growth [17, 19, 20, 25, 28], and low-energy electron diffraction (LEED) [12]. It is known from these studies that the films grow layer-by-layer only below a critical film thickness, which is of the order of 12 unit cells [18]. Characterization by piezoresponse force microscopy (PFM) [29] has demonstrated robust ferroelectricity in films as thin as 1 nanometer [30], while even layers as thin as one unit cell can be ferroelectric in $BaTiO_3/SrTiO_3$ superlattices (1 unit cell = 0.399 nm). However, comprehensive in-situ characterization with surface-sensitive methods to address the atomistic structure, surface termination, and dielectric properties are currently missing, but urgently needed for the optimization of synthesis strategies and to achieve a significant performance increase in magneto-electric structures.

The present paper aims at filling this gap. We present a comprehensive study of the structural, dielectric and electronic properties of $BaTiO_3$ films, which are only a few unit cells thin and grown on Nb-doped $SrTiO_3$ substrates, with a combination of local probe methods and electron diffraction and spectroscopy. We will demonstrate that such films are superior to bulk $BaTiO_3$ substrates regarding the structural quality at the surface, while still showing the dielectric properties of bulk $BaTiO_3$. We propose that $BaTiO_3$ thin films are suitable for fundamental research and applications, such as for the study of magneto-electric effects and magneto-tunnel junctions.

## 2. Experimental section

The $BaTiO_3$ films were grown by molecular beam epitaxy (MBE) by co-evaporation of Ba and Ti from Knudsen cells and using pure ozone as the oxidizing agent. A steady flow of ozone gas was delivered to the growth chamber with the pressure maintained at $2 \times 10^{-6}$ Torr. Prior to film growth, the Nb:$SrTiO_3$ substrates (0.2% Nb doping) were prepared with a buffered HF dip to obtain a $TiO_2$-terminated surface. The substrate was heated in ozone to a growth temperature of 650 °C and then cooled in ozone after film deposition. Each $BaTiO_3$ unit cell was deposited in ~50 s followed by 30 s anneal (with Ba and Ti shutters closed). The deposition was monitored using RHEED.

After growth, the samples were transferred through air into separate ultrahigh vacuum systems for further studies with scanning tunneling microscopy (STM), LEED, photoemission spectroscopy (UPS), etc. While immediately after transfer no LEED pattern was observable, extremely sharp $(1 \times 1)$ diffraction patterns could be established by thermal annealing at approximately 650 K under $O_2$ pressure of $3.75 \times 10^{-7}$ Torr, using an oxygen doser that faces the sample surface at a distance of ~ 5 cm. STM images were taken at 45 K using an Omicron variable temperature scanning tunneling microscope. Photoelectron spectroscopy was carried out at a soft x-ray beam line (3A1) at the Pohang Light Source, Korea. The UPS spectra presented here were taken with the sample kept at room temperature using a hemispherical electron energy analyzer with multichannel detector (Scienta). The energy resolution is 0.3 eV, as determined from the measured shape of Fermi edge of a Cu reference sample. The binding energy is measured with respect to the Fermi edge of Cu. Piezoresponse force microscopy (PFM) measurements were performed in air. An external AC bias voltage is applied to the PFM tip, and the local piezoelectric response from the ferroelectric layer was measured. By scanning the sample surface a two-dimensional map of the piezoresponse amplitude and phase signals could be generated, providing information on polarization magnitude and direction. For local hysteresis loop measurement, a DC offset voltage of controlled magnitude is applied to the tip. Dielectric properties, such as the coercive field and remanent polarization, were deduced from local hysteresis loops.

## 3. Results and discussion

3.1 Surface morphology studies with Scanning tunneling Microscopy

Our studies focus on BaTiO$_3$ films of 8 and 13 unit cells (u.c.) thickness. RHEED images taken along the [100] axis of the pristine Nb:SrTiO$_3$ substrate surface prior to film growth, and of the 8 u.c. and 13 u.c. BaTiO$_3$ films are shown in Figure 1(a)-(c), respectively. Especially for the 8 u.c. film, spots rather than streaks were observed [Figure 1(b)], which is consistent with large, crystalline terraces at the surface of the films. Those spots broaden into streaks as the film thickness is increased to 13 u.c., indicating a gradual increase in surface roughness [Figure 1(c)]. The films were transferred through air into a separate STM UHV chamber after growth, where they were annealed in oxygen to recover ordered LEED diffraction patterns. A weak LEED pattern became observable at an annealing temperature $T_a$ = 150 °C. With increasing temperature the peak intensity and sharpness improved and the background intensity decreased until high-quality (1 × 1) diffraction patterns were observed at $T_a$ ~ 650 °C – 700 °C (Figure 2). Further annealing was found to be detrimental to the diffraction image quality. STM images were taken on both films immediately after the annealing, at a sample temperature of 45 K. Flat terraces of approx. 100 nm width were found on the 8 u.c. film and the terrace step height corresponds to a single unit cell of BaTiO$_3$ (Figure 2). The 13 u.c. films also exhibit atomically flat terraces; however, the terraces are now covered with small islands of average diameter of ~10 nm. A height histogram analysis reveals that there are in average 4 open layers (not shown here). This is also consistent with the observation of streaks in the RHEED images and is attributed to the relaxation of epitaxial film strain. It is concluded that crystalline films showing no detectable surface reconstruction and exhibiting terraces as wide as 100 nm in the case of the 8 u.c. film can be recovered after sample transfer through air.

3.2 Piezoresponse Force Microscopy and Dielectric Properties

Resonance-enhanced PFM measurements [31, 32] have been performed at room temperature on the BaTiO$_3$ films in air after extraction of the samples from the growth chamber, without further sample treatment. Prior to PFM characterization, bi-domain square-in-square polarization patterns have been generated in the BaTiO$_3$ films by scanning the surface with an applied dc bias voltage of ± 4 V. The PFM phase and amplitude maps of the resulting patterns, together with local polarization hysteresis loops, are shown in Figure 3 for the 8 u.c. thin films. Identical results were obtained from the 13 u.c. thin film and are not shown here, for brevity. The amplitude signal in Figure 3(a) is a measure of the polarization magnitude, while the phase signal in (b) shows the polarization direction. The hysteresis loops of the phase and the amplitude signals in (c, d) confirm the existence of non-zero remanent polarization, which can be reversed by applying electric fields larger than those corresponding to the measured coercive bias voltage of approx. 3.5 V. The visible asymmetry in the amplitude signal hysteresis loop for opposite polarization is due to differences in the interfaces on both sides of the $BaTiO_3$. Film boundaries and surface terminations, are known to result in the accumulation of surface charges, which influence the symmetry of the observed PFM loops.

3.3 Photoelectron Spectroscopy and Surface Termination

Photoelectron spectroscopy (PES) data have been collected on the 13 u.c. $BaTiO_3$ film, to determine the surface termination and to learn about the electronic structure of such films. Even though the growth of the films was stopped after deposition of the $TiO_2$ layer, the actual surface termination after transfer through air and annealing in oxygen needs verification. The Ba 4d and the Ti 2p spectra are summarized in Figure 4. The Ti 2p spectrum shows two characteristic peaks identified as the Ti $2p_{3/2}$ and Ti $2p_{1/2}$ peaks. In these spectra, no surface core level shift (SCLS) is observable. Such a shift is often observed at surfaces due to the reduced coordination of the atoms and the resulting change in charge state there. However, to our best knowledge, no SCLS has ever been reported for $TiO_2$-terminated surfaces. The absence of SCLS for the Ti 2p peaks can even be expected since the effective charge of the Ti ions in the $TiO_2$-terminated $BaTiO_3$ surface is similar to that of Ti ions in the bulk $BaTiO_3$, as has been predicted in first principles calculations [10]. Thus, the absence of SCLS of Ti does not exclude $TiO_2$ surface termination.

By contrast, the effective charge of Ba ions in a BaO-terminated surface is less than half of that of Ba ions in the bulk, thereby producing substantial SCLS, as has been reported previously for bulk $BaTiO_3$(001) [33]. For the present films we find the 4d peaks of Ba split into the well-known $4d_{5/2}$ and $4d_{3/2}$ peaks, and two additional peaks shifted by ~1.5 eV towards higher binding energies with respect to those 4d peaks (Fig. 4(b) and (c)). Similarly shifted peaks have been reported for BaO surface layers and ascribed to SCLS [34]. Ba 4d spectra have further been taken with two different photon energies, 695.4 eV and 290.8 eV, shown in Figures 4(b) and (c). We observe an increase of the ratio of surface- to bulk-peaks at lower photon energy where the photoelectrons have smaller kinetic energy and are thus more surface-sensitive. Both observations, the SCLS and the energy dependence of the surface- to bulk peak ratio of the Ba 4d states suggest the existence of BaO in the surface of the films, however, a quantitative analysis of the fractions of BaO- and $TiO_2$-terminated surface areas requires a detailed analysis of the photoemission peak intensities.

The standard approach for the peak intensity analysis in photoelectron spectroscopy [37], has been adapted here for surfaces containing two atomic species. A detailed discussion of the spectral areas of the Ba 4d bulk and SCLS peak pairs is given in the supplementary material. We conclude from this analysis that the BaTiO$_3$ layer is dominantly TiO$_2$-terminated, specifically we find the fractions of the TiO$_2$- and BaO-terminated surface areas to be ~70% and ~30%, respectively. Uncertainties in our analysis result from estimates of the electron mean free path. We performed further complementary experiments with angular-dependent photoemission spectroscopy on several BaTiO$_3$ thin films on SrTiO$_3$ and LaSrMO$_3$ substrates, fabricated by different groups. Those samples have been prepared at varied annealing temperatures (200$^o$C – 650$^o$C) and oxygen partial pressures under UHV. All measurements consistently find surfaces dominantly terminated by TiO$_2$. We conclude that the TiO$_2$ termination of BaTiO$_3$(001) films is energetically favored and robust during in-situ sample preparation.

The valence band spectra of the BaTiO$_3$ samples are shown in Figure 5, together with the spectra of a Cu reference sample for comparison. Most prominently, the valence band, which is mostly of O 2p character, spans from ~3 eV to 10 eV binding energy. The width of the band gap is determined by extrapolating the band edge. This is shown in Figure 5 for two extreme choices of linear extrapolations, resulting in upper and lower limits for the band gap of 2.7 eV – 3.2 eV, respectively. In-gap states are observable between the valence band edge and the Fermi edge. Such states are most commonly found to originate from oxygen defects [33, 35, 36] and Ti 3+ ions located in the vicinity of oxygen defects [36]. While first principles calculations predict that the valence band reaches the Fermi edge [35], we observe instead a band shift to higher binding energy. This shift is similar to the optical band gap of BaTiO$_3$ of 3.22 eV. It is reasonable to attribute this shift to the pinning of the Fermi level at the bottom of the conduction band to the effective *n*-type doping, which is a result of the reduction of the sample during the annealing. This indicates that the film has a band gap analog to bulk BaTiO$_3$, with in-gap states due to local defects. The latter finding is not surprising for the 13 u.c. film given the observed transition to a 3D island structure.

## 4. Conclusions

The advantage of the nanometer thin films of BaTiO$_3$ studied here over bulk samples is the formation of large, atomically smooth surface terraces of unit cell height and the absence of surface reconstruction, while still being insulating and exhibiting remanent electric polarization. Ultrathin BaTiO$_3$ films are thus expected to become substrates for the fabrication of various model systems for the study of magneto-electric effects and also facilitate theoretical analysis. An important result of this work is the demonstration that high-quality surfaces of BaTiO$_3$ can be recovered after sample transfer through air. This will simplify the sample exchange in most lab settings where oxide growth, surface analytics, and device fabrication are done in separate UHV systems. We have evidenced that BaO and TiO$_2$ surface terminations can coexist and are thus similar in energy, thereby providing experimental evidence to a mostly theory-based discussion in the literature. Due to the prevalent $TiO_2$ termination of the surface, these films potentially enable the synthesis of high-quality $Fe/BaTiO_3$ interfaces that might exhibit the recently proposed magneto-electric effect due to Fe-Ti-bond modulation.

**Acknowledgements**

This work has been supported by NSF CAREER (DMR-0747704), NSF MRSEC (DMR-0213808), DoE (DE-SC0004876) and by NRF (2007-0055035). Use of the Center for Nanoscale Materials was supported by the U.S. Department of Energy, Office of Science, Office of Basic Energy Sciences, under Contract No. DE-AC02-06CH11357.

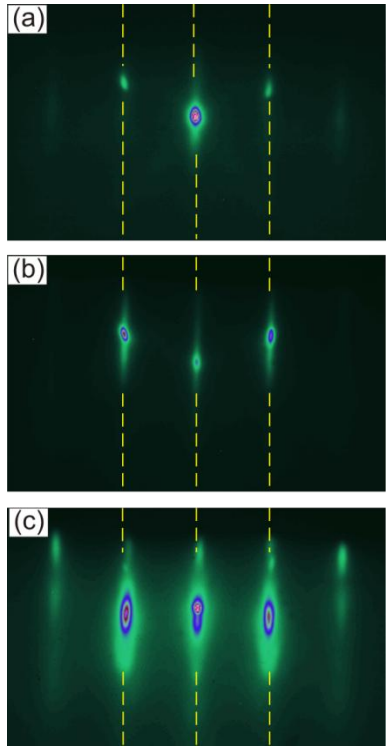

**FIG. 1**: RHEED images at different stages during the growth of BaTiO$_3$ by MBE. (a) pristine Nb-SrTiO$_3$(100) substrate, (b) after growth of 8 unit cells of BaTiO$_3$, (c) after growth of 13 unit cells of BaTiO$_3$ growth.

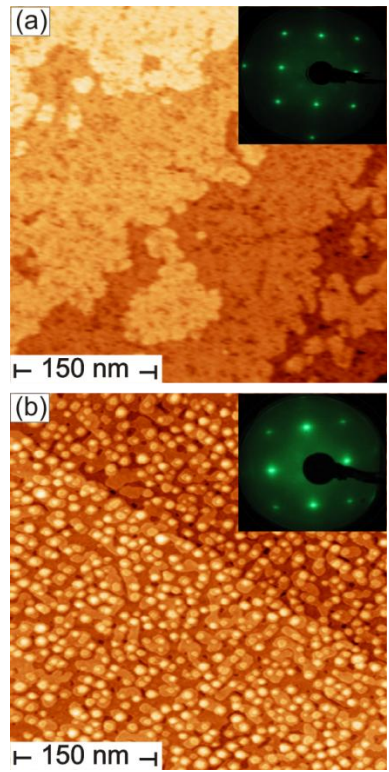

**FIG. 2**: STM images and LEED patterns of BaTiO$_3$ films on Nb-SrTiO$_3$. The thickness of the BaTiO$_3$ films is 8 unit cells (a) and 13 unit cells (b). The LEED images were taken at 85 eV.

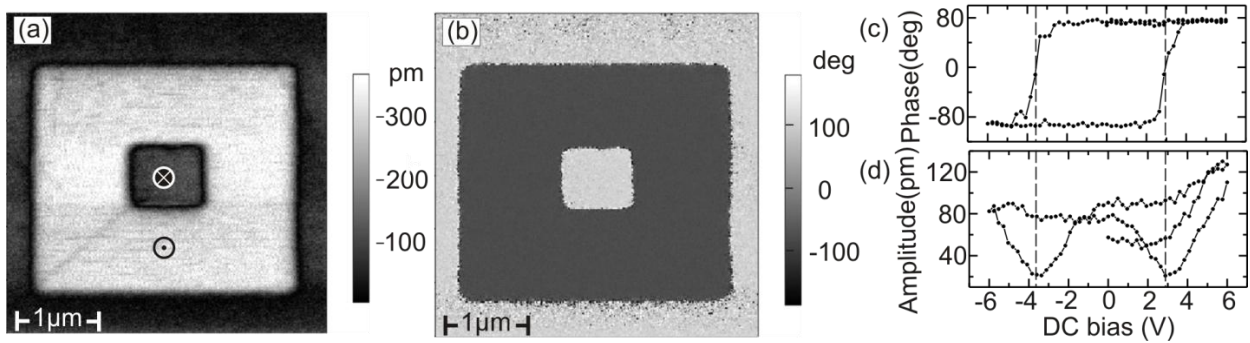

**FIG. 3**: Maps of the remanent piezo-response force microscopy amplitude signal (a) and phase signal (b) of a BaTiO$_3$(8u.c.)/Nb-SrTiO$_3$ film showing areas of opposite ferroelectric polarization, produced by scanning with the tip under ±4V DC bias. Hysteresis loops of the PFM in-field phase (c) and amplitude (d) acquired from the same film.

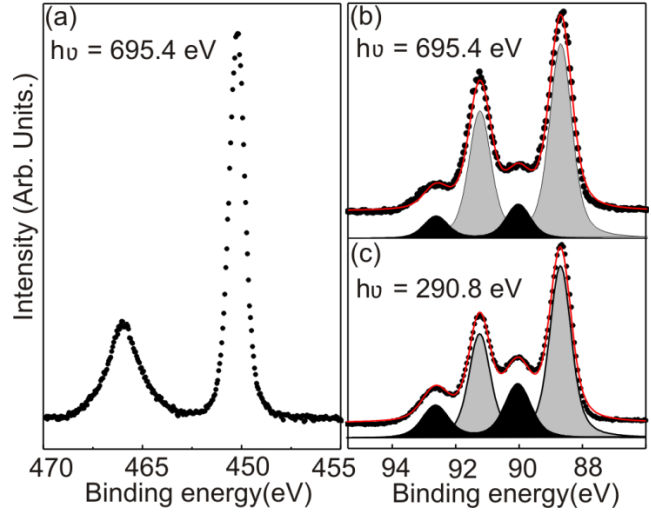

**Fig. 4:** Photoelectron spectra of BaTiO$_3$(001). Ti 2p spectrum (a), Ba 4d spectrum taken at 695.4 eV (b) and 290.8 eV (c), respectively. The binding energy is in reference to Fermi energy. Black- and gray-shaded areas in (b, c) represent fits to the bulk and surface peaks of Ba 4d$_{5/2}$ and 4d$_{3/2}$, respectively.

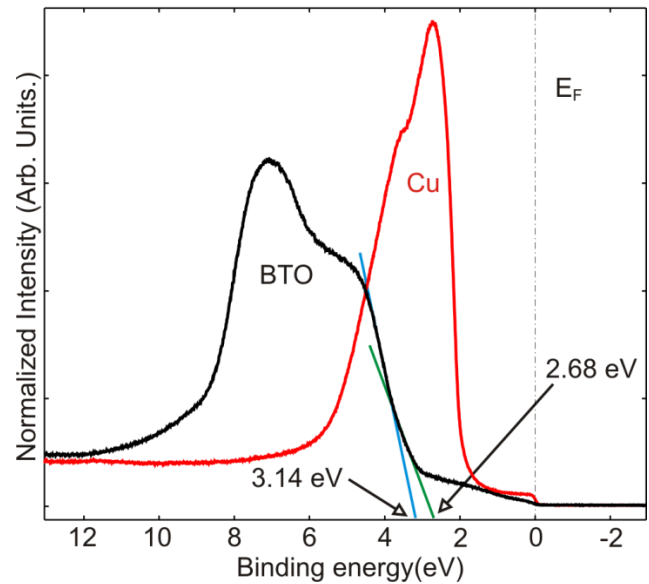

**Fig. 5:** Photoelectron spectra near Fermi edge of clean BaTiO$_3$(001) and Cu. The photon energy was 111.04 eV. Extrapolations of the band edges give the band gap (blue, green lines). The non-zero signal in the band gap is due to the existence of mid-gap states (see text).

**Supporting Information**

The BaO and TiO$_2$ fractions of the surface termination have been obtained from the relative peak intensities of the bulk Ba 4d peak, $I_B$, and the surface Ba 4d peak, $I_s$, following a standard peak intensity ratio analysis found, for instance, in [37]. Considered is a photon flux $F$ incident on the sample surface. The intensity of the 4d peak in the photoelectron spectrum per BaO layer is proportional to the photon flux $F$, the areal density of Ba atoms in the layer $N$, and the photoelectron cross section $\sigma$ for Ba 4d at the given photon energy. We consider now the surface monolayer of the BaTiO$_3$ film, which may consist of a fraction $A$ of BaO, and a fraction $(1 - A)$ of TiO$_2$. We can thus write for the intensity of the surface Ba 4d peak:

$$I_S = C \cdot F \cdot \sigma \cdot A \tag{1}$$

where $C$ is the total area illuminated by the incident photons. For the intensity of the bulk Ba 4d peak, $I_B$, we consider two contributions: one from the BaO-terminated fraction of the film, $I_{B1}$, and one from the TiO$_2$-terminated fraction of the film, $I_{B2}$. Both can be expressed as

$$I_{B1} = C \cdot A \cdot F \cdot \sigma \left( e^{-d/\lambda \cdot \cos\theta} + e^{-2d/\lambda \cdot \cos\theta} + \cdots \right)$$

$$= C \cdot A \cdot F \cdot \sigma \cdot e^{-d/\lambda \cdot \cos\theta} \cdot \frac{1}{1 - e^{-d/\lambda \cdot \cos\theta}} \tag{2}$$

Here, $\theta$ is the angle between the surface normal of the sample and the spectrometer inlet. Considered is the energy- and material-dependent electron escape depth or ineleastic mean free path, $\lambda$, and the interlayer spacing of BaTiO$_3$, $d$. The summation is done as if the film was of infinite thickness, since published values of the escape depth are 4 ~ 7 Å and 2 ~ 4 Å for spectra in Fig. 4(b) and (c), respectively [38]. They are thus much smaller than the film thickness of ~52 Å.

Likewise, we can write for $I_{B2}$:

$$I_{B2} = C \cdot (1 - A) \cdot F \cdot \sigma \left( e^{-d'/\lambda \cdot \cos\theta} + e^{-(d'+d)/\lambda \cdot \cos\theta} + \cdots \right)$$

$$= C \cdot (1 - A) \cdot F \cdot \sigma \cdot e^{-d'/\lambda \cdot \cos\theta} \cdot \frac{1}{1 - e^{-d/\lambda \cdot \cos\theta}} \qquad (3)$$

It is taken into account here that the fraction of TiO$_2$-termination is (1-A) and d' is the interlayer spacing between TiO$_2$ surface layer and the adjacent BaO layer. The fraction of BaO surface termination, *A*, can be determined from the intensity ratio $I_B/I_S$, which is also dependent on the escape depth and interlayer spacing of the BaTiO$_3$:

$$\frac{I_B}{I_S} = \frac{C \cdot F \cdot \sigma \frac{1}{1 - e^{-d/\lambda \cos\theta}} \left[ A \cdot e^{-d/\lambda \cos\theta} + (1-A) \cdot e^{-d'/\lambda \cos\theta} \right]}{C \cdot F \cdot \sigma \cdot A}$$

$$= \frac{1}{1 - e^{-d/\lambda \cos\theta}} \left[ e^{-d/\lambda \cos\theta} + \left(1 - \frac{1}{A}\right) \cdot e^{-d'/\lambda \cos\theta} \right] \qquad (4)$$

It follows for *A*:

$$A = \left[ \frac{I_B}{I_S} \left(1 - e^{-d/\lambda \cdot \cos\theta}\right) + \left(e^{-d'/\lambda \cdot \cos\theta} - e^{-d/\lambda \cdot \cos\theta}\right) \right]^{-1} \qquad (5)$$

Here, we safely assume $d' = d/2$. The ratio $I_B/I_S$ has been measured for 2 different photon energies, and the fitted peak values from the data in Figure 5 are summarized in Table 1. This results in 2 equations to solve for the unknown *A*, assuming for the escape depth 4 ~ 7 Å and 2 ~ 4 Å, respectively in Fig. 5(b) and (c). We thus estimate for the fraction of BaO surface termination, A = 17 − 32%. For the BaTiO$_3$ film, a mechanism for the inelastic scattering, the plasmon loss is missing and the escape depth should be larger than the corresponding one for the metal. Thus, the BaO surface fraction *A* is expected to be closer to ~32 %, which is obtained by inserting the upper limit of the aforementioned escape depths, or 7Å and 4Å, respectively in Fig. 4(b) and (c).

**Table 1.** Fitting parameters and best-fit results of the spectra in Fig. 5(b, c). The parameters GW and LW are Gaussian and Lorentzian widths of the fitting function, respectively.

| Photon energy | 695.4 eV | | 290.8 eV | |
|---|---|---|---|---|
| | Gray | Black | Gray | Black |
| Peak position (eV) | 88.6846 | 90.0207 | 88.6948 | 90.0393 |
| Peak Area | 94341.4063 | 18935.4961 | 63713.1758 | 25481.9961 |
| GW | 0.6250 | 0.7000 | 0.6000 | 0.7500 |
| LW | 0.3500 | 0.4000 | 0.3500 | 0.4500 |
| Spin orbit spitting | 2.55 eV | 2.6 eV | 2.55 eV | 2.6 eV |